\documentclass[USenglish,oneside,twocolumn]{article}
\usepackage[utf8]{inputenc}
\usepackage[big]{dgruyter_NEW}
\cclogo{\includegraphics{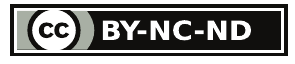}}

\newcommand*\circled[1]{\tikz[baseline=(char.base)]{
            \node[shape=circle,draw,inner sep=1pt] (char) {#1};}}

\usepackage{framed,url,caption,subcaption,graphicx,stringstrings}
\usepackage{amsmath, xcolor,soul,algpseudocode,algorithm,xspace,multirow,longtable,colortbl, lscape}
\usepackage{amssymb, cleveref, todonotes}
\usepackage{booktabs, siunitx,comment}

\definecolor{light-gray}{gray}{0.9}
\definecolor{darkgreen}{rgb}{0,0.5,0}
\definecolor{light-blue}{rgb}{0,.7,1}
\renewcommand{\arraystretch}{1.5}

\graphicspath{{./} }
\crefformat{section}{\S#2#1#3}
\crefformat{subsection}{\S#2#1#3}
\crefformat{subsubsection}{\S#2#1#3}

\expandafter\def\expandafter\UrlBreaks\expandafter{\UrlBreaks
  \do\a\do\b\do\c\do\d\do\e\do\f\do\g\do\h\do\i\do\j%
  \do\k\do\l\do\m\do\n\do\o\do\p\do\q\do\r\do\s\do\t%
  \do\u\do\v\do\w\do\x\do\y\do\z\do\A\do\B\do\C\do\D%
  \do\E\do\F\do\G\do\H\do\I\do\J\do\K\do\L\do\M\do\N%
  \do\O\do\P\do\Q\do\R\do\S\do\T\do\U\do\V\do\W\do\X%
  \do\Y\do\Z}

\newcount\Comments  \Comments=1  
\newcommand{\kibitz}[2]{\ifnum\Comments=1\textcolor{#1}{#2}\fi}

\newcommand{\R}{\mathbb{R}}

\newcommand{\paratitle}[1]{\vspace{.05in} \noindent\textbf{#1}}

\providecommand{\ie}{\emph{i.e.,} }
\providecommand{\eg}{\emph{e.g.,} }

\providecommand{\etal}{\emph{et al.\xspace}}
\providecommand{\etc}{\emph{etc.\xspace}}

\newcommand{\boldred}[1]{\textcolor{red}{\textbf{#1}}}

\newcommand{\presec}{\vspace{-0.15in}}
\newcommand{\postsec}{\vspace{-0.1in}}
\newcommand{\presub}{\vspace{-0.2in}}
\newcommand{\postsub}{\vspace{-0.1in}}
\newcommand{\presubsub}{\vspace{-0.15in}}
\newcommand{\postsubsub}{\vspace{-0.1in}}

\newcommand{\precaption}{\vspace{-0.2in}}
\newcommand{\postcaption}{\vspace{-0.2in}}

\newcommand{\name}{\textsc{Kashf}\xspace}

 \author*[1]{John Cook}

  \author[2]{Rishab Nithyanand}

  \author[3]{Zubair Shafiq}

  \affil[1]{The University of Iowa, E-mail: john-cook@uiowa.edu}

  \affil[2]{The University of Iowa, E-mail: rishab-nithyanand@uiowa.edu}

  \affil[3]{The University of Iowa, E-mail: zubair-shafiq@uiowa.edu}

\begin{document}

\title{\huge Inferring Tracker-Advertiser Relationships \\in the Online Advertising Ecosystem using Header Bidding}

\runningtitle{Inferring Tracker-Advertiser Relationships in the Online Advertising Ecosystem using Header Bidding}

\begin{abstract}{
Online advertising relies on trackers and data brokers to show targeted ads to users.
To improve targeting, different entities in the intricately interwoven online advertising and tracking ecosystems are incentivized to share information with each other through client-side or server-side mechanisms. 
Inferring data sharing between entities, especially when it happens at the server-side, is an important and challenging research problem.
In this paper, we introduce \name: a novel method to infer data sharing relationships between advertisers and trackers by studying how an advertiser's bidding behavior changes as we manipulate the presence of trackers. 
We operationalize this insight by training an interpretable machine learning model that uses the presence of trackers as features to predict the bidding behavior of an advertiser. 
By analyzing the machine learning model, we can infer relationships between advertisers and trackers irrespective of whether data sharing occurs at the client-side or the server-side.
We are able to identify several server-side data sharing relationships that are validated externally but are not detected by client-side cookie syncing.
}\end{abstract}


	\journalname{Proceedings on Privacy Enhancing Technologies}
	\DOI{10.2478/popets-2020-0001}
	\startpage{1}
 	\received{2019-05-31}
	\revised{2019-09-15}
	\accepted{2019-06-16}

	\journalyear{}
	\journalvolume{2020}
	\journalissue{1}
	
\maketitle

\presec
\presec
\presec
\section{Introduction} 
\label{sec:intro}
\postsec

\paratitle{Online vs. offline advertising.} 
Online advertising is set to surpass offline advertising (\eg newspapers, yellow pages, radio, TV) this year.
In fact, online advertising revenues in the US are expected to exceed
two-thirds of total advertising spending by 2023 \cite{emarketer19USadspending}. 
There are several reasons driving this shift from offline advertising to online advertising. 
First, consumers are increasingly spending more time online. 
This makes the web a more attractive platform for advertisers. 
Specifically, consumers in the US now spend about 24 hours a week online, which exceeds the time spent watching TV \cite{digitalfuture17,molla18onlinevstv}.
Second, online advertising primarily relies on highly automated technologies that enable advertisers to programmatically launch advertising campaigns, measure their effectiveness, and quickly adjust them based on their performance. 
Programmatic advertising already accounts for 86\% of all online display advertising in the US
\cite{fisher18usprogrammaticadspend}.
Third, online advertising allows targeting of advertising campaigns to specific audiences based on their demographics, location, or intents. 
Personalized online advertising campaigns are reported to be much more effective as compared to their non-personalized counterparts \cite{jivox16personalization}.

\paratitle{The online advertising ecosystem includes middle-men.} 
Unlike offline advertising where there is typically a direct relationship
between advertisers and publishers, the online advertising ecosystem comprises
of several specialized entities that mediate interactions between advertisers and publishers. 
This is necessitated by the need for technical expertise to
participate in protocols, such as real-time bidding (RTB), which require
publishers and advertisers to identify, offer, and respond to ad impression
opportunities in near real-time.
The entities that fill this gap include \textit{supply-side platforms} (SSPs) that put up ad inventory of  publishers for sale at \textit{ad exchanges} (AdXes), which are marketplaces that run real-time auctions for individual ad slots. 
Advertisers bid on individual ad slots auctioned off at AdXes through
\textit{demand-side platforms} (DSPs), which use sophisticated models to
determine how much to bid  for an ad slot based on the user information
retrieved from \textit{data management platforms} (DMPs). 
DMPs gather user information (\eg browsing history) through a variety of online tracking techniques such as cookies.  

\paratitle{Entities engage in data sharing.}
Intuitively, an ad slot's value as assessed by a DSP, is highly dependent on the quality of information received from the DMP(s). 
Consequently, DMPs strive to enhance the quality of information by improving their ability to observe user behavior on different websites and platforms. 
This can be done by either (1) increasing their presence, as trackers, on the web or (2) developing data sharing relationships with other tracking services. 
Prior research has shown that only a few organizations (Google, Facebook, Twitter, Amazon, AdNexus, and Oracle) are able to track users on more than 10\% of the top 1-million sites \cite{Englehardt16CCSOWPM}. 
Thus, DMPs often choose to develop data sharing relationships rather than trying to arduously increase their presence on the web. 
In fact, the RTB protocol has built-in mechanisms to facilitate data sharing between advertisers and trackers. 
Cookie syncing (a.k.a. cookie matching) in RTB allows two different entities in RTB to exchange their cookies while bypassing the same-origin policy \cite{OLEJNIK14NDSSRTB,ACAR2014CCS}. 
Cookie syncing essentially allows two entities to map their cookies to each other and get a more complete view of a user's browsing history \cite{Ghosh15tomatchornottomatch}.
A recent study showed that cookie syncing increases the number of entities that track users by almost 7X \cite{Papadopoulos19cookiesynchronization}. 
Another recent study showed that, despite using privacy-enhancing technologies such as Ghostery and Disconnect, trackers are still able to observe anywhere from 40-80\% of a user's browsing history due to cookie syncing in RTB \cite{Bashir18PETSDIFUSSION}.

\paratitle{Transparency of data sharing relationships is important.}
Privacy researchers and regulators are increasingly interested in studying data sharing
relationships between different entities in the intricately interwoven online advertising
and tracking ecosystems for several reasons. 
First, a complete understanding of such relationships can help detect whether a domain is a tracker and, in turn, improve the effectiveness of tracker blocking tools \cite{kalavri16communitywebtrackers}. 
Blocking tools are presently the most effective protection users can employ against trackers.  
Second, it is important to uncover data sharing relationships between different organizations for regulatory compliance verification purposes. 
Both General Data Protection Regulation (GDPR) \cite{URL_GDPR} in Europe and the California Consumer Privacy Act  (CCPA) \cite{URL_CCPA} in the US give people the right to know what personal information is being collected and whether (and with whom) it is being shared. 
Methods that can detect data sharing between different tracking/advertising
organizations can help uncover unauthorized or undisclosed data sharing relationships.

\paratitle{Measuring client-facilitated data sharing is insufficient.}
Analysis of client-facilitated mechanisms in RTB, such as cookie syncing, to detect data sharing between entities is limited due to two reasons.  
First, prior research relies on different heuristics to detect cookie syncing at the client-side \cite{Papadopoulos19cookiesynchronization,OLEJNIK14NDSSRTB, ACAR2014CCS}. 
Unfortunately, these heuristics are brittle to changes in non-standardized implementations of cookie syncing, especially when obfuscation is employed \cite{Bashir16USENIXTRACING}.  
Second, and more importantly, analysis of client-side mechanisms such as cookie syncing \emph{cannot detect server-side data sharing between entities} \cite{sst_woopra}. 
Server-side tracking (\eg postback tracking \cite{URL_POSTBACK}) is expected to grow in popularity as mainstream browsers, notably Safari and Firefox \cite{URL_ITP_20,URL_FF_ETP}, have started to implement stringent third-party cookie policies \cite{itp20_PartnerStack}. Thus, it is important to develop methods that can infer both client-side and server-side data sharing between different entities in the online advertising ecosystem. 

\paratitle{Inferring server-side data sharing is challenging.}
It is particularly challenging to infer server-side data sharing
because it is not directly observable from purely client-side measurements. To
overcome this challenge, prior research has attempted to exploit artifacts that
reflect semantics of how online ads are served, rather than relying on specific mechanisms such as cookie syncing. In a seminal work, Bashir et al.
\cite{Bashir16USENIXTRACING} exploited retargeting to infer data sharing
even if it occurs on the server-side. Their key insight is that retargeting
takes place only when data sharing occurs between AdXes on different
sites. To operationalize this insight, the authors trained personas to trigger
retargeting, which is detected using crowdsourcing, and then analyze inclusion chains to determine whether information is shared at client-side or server-side. 
Using retargeting to infer server-side data sharing is limited because retargeting represents only a subset of scenarios in which server-side data sharing occurs.
More specifically, server-side information exchange is a necessary but not a sufficient condition to trigger retargeting.
Furthermore, detecting retargeting is a challenging task that requires significant manual effort that
is not only difficult to scale but also susceptible to human errors. 

\paratitle{Inferring tracker-advertiser data sharing using header bidding.} 
To address our inability to directly measure server-side data sharing, like previous work \cite{Bashir16USENIXTRACING}, we also leverage client-side observable artifacts of the online advertising ecosystem.
However, instead of relying on retargeting, we rely on being able to observe the bids placed by DSPs or bidders (on behalf of advertisers) that participate in \textit{header bidding} (HB) -- a new programmatic advertising mechanism aimed at increasing publisher advertising revenue as compared to traditional RTB. 
In contrast to traditional RTB that only exposes the winning bid at the client-side, HB exposes \emph{all} bids made by different advertisers at the client-side.
The ability to precisely observe the bids placed by a given advertiser in HB\footnote{It is noteworthy that in RTB only exposes the winning bid and the corresponding winner in the auction. Moreover, even when a given advertiser wins the auction, RTB does not expose the highest bid due to its use of the second-price auction. Thus, RTB does not allow us to observe the bid placed by a given advertiser.} enables us to observe how advertiser bids vary as a persona's browsing history and tracker presence are modified. 
At a high-level, our approach (named \name) to inferring tracker-advertiser data sharing relies on the following insight: \emph{advertisers with knowledge of a user's browsing history will bid differently (potentially higher) than an advertiser having no knowledge of a user's browsing history}. 
In order to operationalize this insight, we first selectively expose a persona's browsing history to different sets of trackers and record the bids made by an advertiser in HB. 
We then train interpretable machine learning models, using tracker presence as input features and bid values as the target variable, to accurately predict the bids made by an advertiser.
We finally leverage the interpretability of the trained machine learning models to determine which features (\ie trackers) were most influential in predicting the values of bids placed by an advertiser.
This enables us to make inferences about the presence of data sharing relationship (client-side or server-side) between advertisers and trackers.

\paratitle{Key contributions.} This paper makes the following two key contributions.

\begin{itemize}
    \item \textit{Measuring bidding behavior of advertisers} (\Cref{sec:quantifying}). 
        We are able to draw several novel insights into how different advertisers value users. 
        More specifically, we leverage HB, which exposes \emph{all} bids made by different advertisers, to study how an advertiser's bidding preferences vary for different personas. 
        We find that with the exception of users with a \emph{Health} persona who are universally preferred, advertisers have very different persona preferences. 
        We also find that advertisers often have a strong preference for certain personas and these rarely overlap. 
        Furthermore, we are able to shed light on the practice of underbidding. 
        We find that underbidding is very common with zero bids making up 22\% of all bids. 
        
    \item \textit{Uncovering tracker-advertiser relationships} (\Cref{sec:inferring}).
        Our approach, \name, allows us to infer data sharing relationships between advertisers and trackers irrespective of whether they occur at the client-side or the server-side.
        To this end, we train machine learning models that use tracker information as features to predict the bidding behavior of advertisers with 75-83\% accuracy.
        By analyzing the interpretable machine learning model, we are able to identify data sharing relationships between advertisers and popular trackers, most of which we are able to externally validate.
        We also demonstrate that many of these inferred server-side data sharing relationships are not detected by client-side cookie syncing.
\end{itemize}

\presec
\presec
\section{Background} \label{sec:background} 
\postsec

In this section, we provide an overview of online advertising and tracking
ecosystems and highlight how they are intertwined.  

\begin{figure*}[!t] 
  \centering 
  \includegraphics[width=1.0 \textwidth]{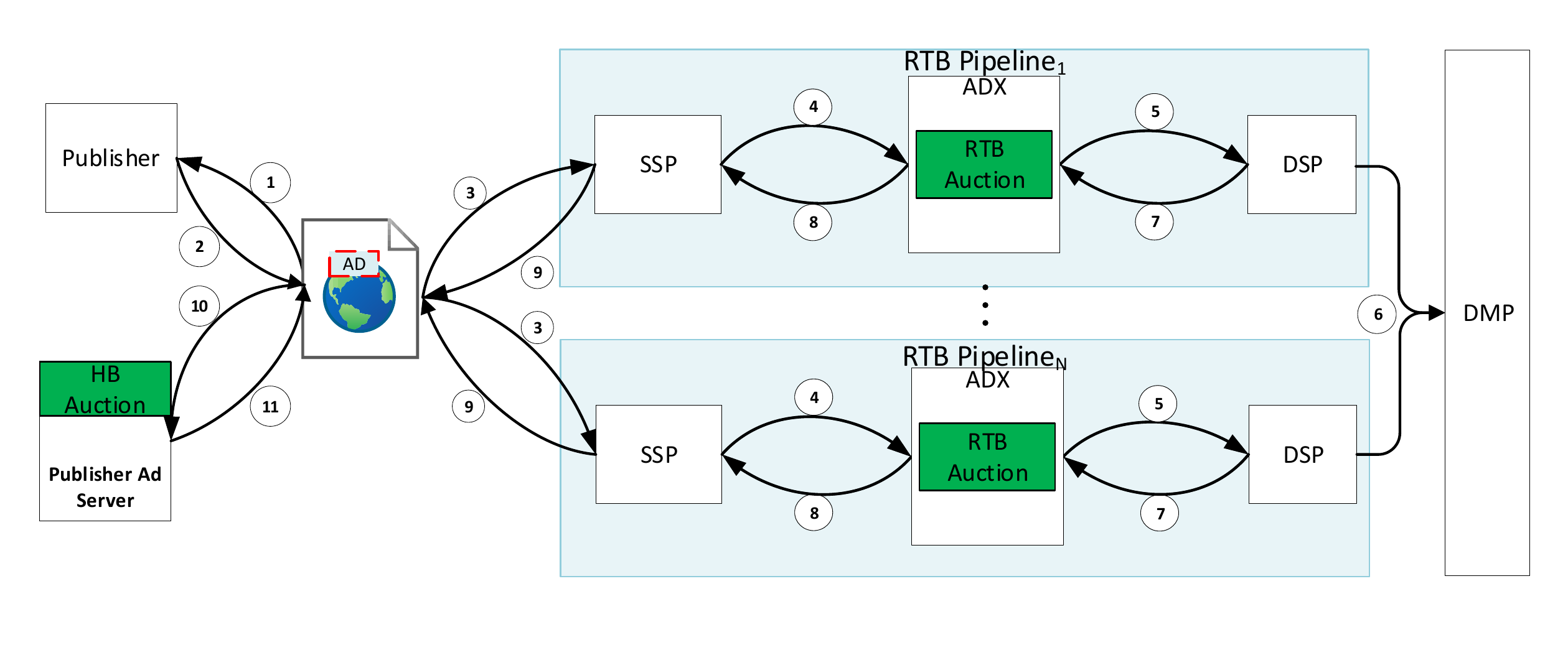}
\precaption
\precaption
  \caption{Online advertising workflow using Real-Time Bidding (RTB) and Header Bidding (HB).} 
\label{fig:RTBHB} 
\postcaption
\end{figure*} 

\presub
\subsection{Online Advertising Ecosystem} \label{sub:adecosystem} 
\postsub

The contemporary online advertising ecosystem relies on programmatic processes to trade ad impressions in near real-time (\ie typically less than 100ms).

\paratitle{Real-Time Bidding (RTB).} RTB is the most widely used programmatic process in online advertising. 
The typical RTB workflow illustrated in \Cref{fig:RTBHB} involves interactions between several entities in the advertising ecosystem. 
These include publishers, publisher ad servers, supply side platforms (SSPs), ad exchanges (AdX), demand side
platforms (DSPs), and data management platforms (DMPs). 
The RTB process has three distinct phases: ad request, bid collection, and ad placement.

\begin{itemize}

  \item \textbf{Ad request.} The workflow is initiated when the browser sends
    a request to fetch the publisher's web page \circled{1}. 
    The publisher's web server responds with the HTML document that contains page content as
    well as the ad tag \circled{2}.
    While the rest of the page is loaded, the ad tag generates an ad request to an RTB-enabled SSP along with information about the ad slot (e.g., dimension, media type) \circled{3}. 
  \item \textbf{Bid collection.} The SSP's role is to manage the
    publisher's ad inventory and put it up for auction at an AdX \circled{4}.
    The AdX notifies the DSP of the available ad inventory by sending a bid
    request \circled{5}, which is composed of information from the ad slot as
    well as any identifiers from the browser (\eg via cookie syncing
    \cite{OLEJNIK14NDSSRTB,URL_COOKIEMATCHING}).
    The DSP evaluates the bid request using the information sent by the AdX and
    by synchronizing any identifiers with one or more DMPs \circled{6}. 
    The DSP then acts on behalf of an advertiser by generating and sending the AdX
    a bid \circled{7}. 
    DSPs typically implement sophisticated bidding strategies
    that leverage campaign information from the advertiser, tracking
    information from the DMP(s), and ad slot information in the bid request. 
  \item \textbf{Ad placement.} The AdX collects bid responses from multiple
    DSPs and uses an auction mechanism (typically a second price auction) to
    determine the winning bid. 
    If the winning bid value surpasses the impression's minimum sale price set by the publisher, the winning bid and the associated ad is forwarded to the SSP \circled{8}, which places the ad on a browser page
    \circled{9}. 
    If the winning bid value does not surpass the impression's minimum sale price, the impression is presented to the next preferred AdX (as determined by the SSP) and the bidding process is repeated.
    Auctions occurring at lower levels of the ``waterfall'' have a residual effect on bidder perception, resulting in progressively lower bids. 
\end{itemize}

\paratitle{Header Bidding (HB).} HB is an emerging programmatic process for online
advertising that is rapidly gaining popularity due to its promise to
increase yield for publishers as compared to traditional RTB
\cite{URL_HBVSRTB,Pachilakis19imc}.
According to a recent survey \cite{URL_HB2018}, more than half of the top one thousand websites that offer programmatic advertising already use HB. 
In contrast to RTB, where the ad inventory is offered to different ad exchanges
(and consequently bidders) in a sequential (or waterfall) manner, HB offers
the ad inventory to multiple bidders simultaneously. 
More specifically, in the waterfall model used by RTB, the ad inventory is first offered to higher tier
ad exchanges and any leftover inventory is offered to lower tier exchanges.
This sequential process results in less competition for bids and
subsequently reduces publisher yield from advertising. 
HB essentially flattens the waterfall, forcing increased competition among different bidders for ad
impressions and increasing the yield for publishers.
While there is some overlap between RTB and HB, we explain the workflow of the
HB model by discussing differences that occur in the ad request (steps
\circled{2} and \circled{3}) and ad placement (\circled{9} -- \circled{11})
phases. 
The bid collection process (\circled{4} -- \circled{8}) remains
unchanged in HB.

\begin{itemize}
  \item \textbf{Ad request.} In HB, the publisher's web server responds with
    the HTML document that contains page content as well as the ad tag with
    a HB \textit{wrapper}.\footnote{There are two common implementations of HB.
    We are discussing client-side HB as opposed to server-to-server HB
    \cite{URL_S2SAPPNEXUS}.}
    The HB wrapper pauses a page's ad tag from being executed and sets
    a predetermined timeout. 
    While the ad tag is paused, the wrapper simultaneously contacts different demand partners (mainly SSPs) by sending them bid requests \circled{3}. 
    While the HB wrapper is awaiting bid responses, parallel auctions are occurring in multiple RTB pipelines as shown in Figure \ref{fig:RTBHB}.
  \item \textbf{Ad placement.} Each SSP asynchronously sends bid responses to the HB wrapper \circled{9}. 
    Once the HB timeout expires, all bids are then forwarded to the publisher's ad server \circled{10} where a unified HB auction mechanism is used to determine the winning bid and price. 
    The browser is notified of the winning bid and the corresponding ad is placed
    on the page \circled{11}.
\end{itemize}

\presub
\subsection{Online Tracking Ecosystem} \label{sub:trackingecosystem}
\postsub

The online tracking ecosystem is composed of a large number of organizations
engaging in tracking user behavior across the web. 
This is accomplished by a variety of techniques including tracking cookies, pixel tags, beacons, and
other sophisticated mechanisms. 
Below we provide a high-level overview of how online tracking works and some aspects of the interplay between online tracking and online advertising.

\paratitle{How online tracking works.} In order to deterministically identify
users across the web, trackers need to assign unique identifiers to individual
users. 
This is often accomplished using cookies. 
Cookies are stored at the client-side and are typically structured as key-value pairs that
contain identifiers that uniquely identify a user.
A tracker present, as a third-party, on multiple domains across the web can read their own cookies linked to a user as they traverse domains. 
This enables individual trackers to recreate subsets of a user's browsing history.

There are two key limitations of cookies from the perspective of online
tracking. 
First, due to the {same-origin policy} enforced by browsers, access
to a cookie is restricted to the tracker that sets it. 
This means that two trackers, each with a partial view of a user's browsing history, cannot enhance
their knowledge by directly sharing their own cookies with each other. 
To circumvent this limitation, trackers typically rely on \emph{cookie
syncing} in order to map each other's identifiers of a user \cite{OLEJNIK14NDSSRTB}. 
Second, since they are stored at the client-side (browser), {cookies can be deleted by users}.
While trackers can always set new cookies, there is still no sound way of
linking deleted cookies with new cookies. 
To circumvent this limitation, trackers now also rely on \emph{cookie respawning}
\cite{ACAR2014CCS} and other stateless (probabilistic) techniques such as \emph{browser
fingerprinting} \cite{Englehardt16CCSOWPM}.

\paratitle{Tracker relationships.} In order to generate a more complete view of
a user's browsing history and interests, trackers may collaborate with one
another. 
This is accomplished by using client-side or server-side mechanisms to share
information. 
Client-side mechanisms rely on the user's browser to facilitate an information sharing channel between the collaborating trackers. 
As a result, these data sharing relationships are directly observable at the client-side. 
Cookie syncing is a popular client-side mechanism that is used by trackers to
facilitate cookie sharing between trackers even in the presence of the
same-origin policy. 
Other client-side mechanisms involve the sharing of other
(non-cookie) unique identifiers such as email addresses and unique device
identifiers (\eg IMEI, Android ID, \etc) \cite{Vallina-RodriguezNarseo2016TtTT, Fouad19trackingpixels}.  
Server-side mechanisms may rely on an out-of-band information sharing channel
between collaborating trackers. 
Since the user's browser is not involved in the mechanism, these data sharing relationships are not directly observable at the client. 
Instead, more complex controlled experiments are needed to infer such
relationships \cite{Bashir16USENIXTRACING,Bashir18PETSDIFUSSION}.

\paratitle{Synergy between online advertising and tracking.} Two common
strategies used by online advertisers for targeting users are contextual
targeting and behavioral targeting. In contextual targeting, ads shown to
a user are only dependent on the content of the page (or website) being viewed.
In contrast, behavioral advertising shows ads that are based on the interests
and behavior demonstrated by the user. In recent years, advertisers have
started to increasingly rely on behavioral advertising. In fact, just between
2008 and 2018, the ad spend on behavioral ads in the United States
increased from \$775M to \$47B \cite{URL_HB2018}. 
While many in the online advertising industry claim that behavioral advertising is always more effective that contextual advertising, its effectiveness relies on the
quantity and quality of data available about the targeted individual. 
As a direct consequence, user data obtained from online trackers is vital to the
success of the advertising industry. Put another way, data obtained by online
tracking is deemed critical to improving the click-through rate (CTR) and return on
investment (ROI) in online advertising campaigns
\cite{URL_ECONOMICCONTRIBUTIONSEUROPE}.
Data management platforms (DMPs), shown in \Cref{fig:RTBHB}, are responsible
for feeding user data obtained by trackers into the advertising ecosystem
bidding process. 

\presec
\section{Quantifying the Value of Users} \label{sec:quantifying}
\postsec
In this section, we seek to understand how much advertisers are \textit{willing to pay} to reach different users. 
Prior work has attempted to understand how much advertisers \textit{pay} to reach different users \cite{OLEJNIK14NDSSRTB, PAPADOPOULOS17IMCRTB, GONZALEZACMCHI17CHIFDTV}. 
There is a subtle but important difference between our and prior work. 
Prior work is limited to studying the price \textit{actually paid} by only the winning bidder in RTB. 
Specifically, prior work leveraged the winning price notifications in RTB, which only exposes the winning bid at the client-side. 
First, note that the winning price is actually \textit{not} the bid value of the winner but rather the bid value of the second-highest bidder plus a predetermined amount (typically one cent) because RTB uses the second-price auction.
Second, note that RTB's winning price notification does not include any information about the bidders (or their bids) that did not win the auction. 
Third, note that RTB's winning price notifications often encrypt the winning bid, which prior work \cite{OLEJNIK14NDSSRTB} assumed (incorrectly \cite{PAPADOPOULOS17IMCRTB}) to be the same as plaintext bids. 
In this section, we leverage HB to empirically understand how much advertisers are \textit{willing to pay} to reach different users while avoiding the limitations of prior work based on RTB.
The HB process, as explained in \Cref{sub:adecosystem}, typically requires that \emph{all} bids made by different bidders are forwarded to the publisher ad server via the client in plaintext. 
This client-side access to the details of every bid made by advertisers for different user personas, not just winning bids, allows us to improve and extend the analytical insights drawn by previous work \cite{OLEJNIK14NDSSRTB, PAPADOPOULOS17IMCRTB, GONZALEZACMCHI17CHIFDTV}. 
\Cref{tab:quantify:questions} illustrates the contributions of our work towards quantifying the value of users to advertisers.
\begin{table}[!t]
  \centering
  \small
  \begin{tabular}{p{2.75in}c}
    Question & Results \\
    \midrule
    How much does a user's persona impact bid values?
    & \Cref{sub:quantify:persona} \\
    How much does user intent matter to advertisers?
    & \Cref{sub:quantify:intent} \\
    \textbf{How does bidding behavior vary across advertisers?}
    & \Cref{sub:quantify:advertisers} \\
    {How much do advertisers pay to reach users?} & \Cref{sub:quantify:winning_bids} \\
    \textbf{How common is underbidding?} & \Cref{sub:quantify:underbidding} \\
  \end{tabular}
  \caption{Questions answered by our study. Questions in \textbf{bold} have not
  been answered by previous work.}
  \label{tab:quantify:questions}
  \postcaption
\end{table}

\presub
\subsection{Measurement Method} \label{sec:quantifying:methodology}
\postsub

To answer the questions listed in \Cref{tab:quantify:questions}, we conducted controlled experiments using the following method. 
At a high-level, our measurement method is explained by: 
(1) how we crawl web pages, 
(2) how we create web personas which can signal user intent to complete a transaction, 
(3) how we gather the bids placed by HB participants on HB-enabled websites.

\paratitle{Web crawling.} 
Our measurements were conducted using a lightly modified version of OpenWPM \cite{Englehardt16CCSOWPM}. 
OpenWPM was used to automatically load selected webpages. 
The timeout for each page load was set to 60 seconds. 
In order to more accurately simulate real user behavior, the bot-mitigation features of OpenWPM were enabled and additional scrolling on the webpage was performed 5 seconds after the browser on-load event fired.
Randomized delays of 2-7 seconds were implemented between each page scroll.

\paratitle{Creating web personas.} 
In order to gather information about how advertisers value different users, we need to construct \emph{personas} mimicking different users. 
We constructed personas based on each of the 16 categories found on Alexa's top sites by category \cite{URLALEXASITESBYCATEGORY}. 
Starting with a clean slate (clean client state), we crawled the top 50 sites in each of these categories using the OpenWPM configuration described above, saving associated browser cookies after each site visit. 
Each web persona is an accumulation of cookies for a single category and is fully constructed when the crawl is complete.
No crawling is performed to construct the control persona -- its persona is the absence of cookies. 
Note that our approach to constructing web personas aligns with previous work within this space \cite{OLEJNIK14NDSSRTB,Bashir16USENIXTRACING}.
\Cref{tab:quantify:personas} lists the 16 different personas used in our
study.

\begin{table}[!t]
  \centering
  \small
  \begin{tabular}{p{.5in} p{2.7in}}
      Personas created & Adult, Art, Business, Computers, Games, Health, Home,
      Kids, News, Recreation, Reference, Regional, Science, Shopping,
      Society, Sports.\\\hline
      Intent sites & \texttt{hotels.com}, \texttt{zales.com},
      \texttt{jamesedition.com}, and \texttt{luxuryrealestate.com}.\\
  \end{tabular}
  \caption{List of web personas created and sites used to convey transaction
  intent by our study. Each persona reflects a user browsing the most popular
  sites in the corresponding Alexa category. A product page was browsed on each
  intent site to convey transaction intent.}
  \label{tab:quantify:personas}
  \postcaption
\end{table}

\paratitle{Signaling intent.} Prior work \cite{OLEJNIK14NDSSRTB} showed that advertiser bids were generally higher for personas which had previously demonstrated intent to make a transaction (\eg by navigating to a specific product page on a website). 
We follow the methodology in \cite{OLEJNIK14NDSSRTB} and select a small number of sites to signal intent. 
\Cref{tab:quantify:personas} lists the 4 sites on which specific products were chosen to demonstrate transaction intent. 
In order to create personas which signal transaction intent, we repeated the persona construction method detailed above followed by the intent signaling mechanism described here. 
We constructed \emph{intent} and \emph{no intent} versions of each of our 16 personas.

\paratitle{Gathering advertiser bids.} 
After constructing intent/non-intent personas, we crawled HB-enabled websites to gather advertiser bids for each of our personas. 
In order to identify HB-enabled websites, we crawled the Alexa top 10K websites and shortlisted the 25 most popular domains (e.g., espn.com, accuweather.com, cnn.com) that support the most well-known open-source implementation of HB called \texttt{prebid.js} \cite{URL_ADZERK, URL_ADEXCHANGER}. 
This was done by checking the \texttt{prebid.js version} attribute in the \texttt{prebid.js} client-side API.\footnote{Our measurements are dependent on client-side implementation of {\tt prebid.js}, which would not work for server-side  implementation \cite{URL_S2SAPPNEXUS,Pachilakis19imc}.}
Domains returning valid responses were marked as HB-enabled.
During each visit to these 25 websites, we made a call to the \texttt{prebid.js} API's \texttt{getBidResponses} method. 
All bid responses returned by this method reflect the bids placed by advertisers for the persona/intent being tested.
The bid responses were saved and formed the basis of our analysis. 
We repeated this process 10 times for each intent/no-intent persona.

\subsection{Results}\label{sub:quantify:results}
\postsub
We now analyze the bids placed by advertisers for the following 33 personas: 16 personas signaling no intent to complete a transaction, 16 personas signaling intent to complete a transaction, and one control persona with no prior browsing history. 
Next, we use these bids to answer the questions listed in \Cref{tab:quantify:questions}.

\presubsub
\subsubsection{How much does a user's persona impact bid values?}
\label{sub:quantify:persona}
\postsubsub

\Cref{tab:quantify:nointent} shows the median bid values by the five most prevalent bidders for each of our 16 personas and the control persona. 
The bid values are expressed in \emph{cost-per-mille} (CPM) which reflects the price paid for 1,000 ad impressions.
Focusing on the average column, we are able to draw several conclusions about the impact of personas on average bid values.
First, we note that the \emph{Control} persona (without any browsing history) attracts lower bids than most of the trained personas.
Second, we note that the \emph{Health} persona attracts significantly higher bids -- 1.6x the average bid value across all categories. 
Similarly, bids are significantly above the average across all categories for \emph{Computers}, \emph{Science}, and \emph{Shopping} personas. 
Third, we note the \emph{Sports} persona attracts the lowest average bids -- 0.6x the average bid value across all categories.
Similarly, bidders bid significantly below the average across all categories for \emph{Games}, and \emph{Home} personas. 
Finally, we note significantly high variation in bid values across bidders for the \emph{Health} and \emph{Computer} personas, which also receive significantly higher bids than other personas.  
Overall, variability in average bid values allows us to conclude that \emph{a user's persona impacts bids placed by an advertiser.}

\presubsub
\subsubsection{How much does user intent matter to advertisers?}
\label{sub:quantify:intent} 
\postsubsub

\Cref{tab:quantify:intent} shows the ratio of the median bids placed by the five most common bidders for personas showing intent to those showing no intent. 
A ratio near 1 means that bid prices remain similar between \emph{Intent} and \emph{No-Intent} personas.  
Focusing again on the Avg. column, we are able to draw several conclusions about the impact of showing intent on average bid values. 
First, we note that the \emph{Control} persona (without any browsing history) attracts significantly lower bid ratios than most of the trained personas.
It is also notable that bid ratios from the \emph{Control} persona were near 1 -- bidder express significantly lower interest regardless of intent. 
Second, we note that the \emph{Health} persona attracts the highest average bid ratios -- 2.4x higher than the \emph{No-Intent} \emph{Health} persona. 
A notable point is that the \emph{Health} persona attracts the highest bid values (\Cref{tab:quantify:nointent}) and bid ratios (\Cref{tab:quantify:intent}) -- on average bidders are willing to pay significantly higher prices for \emph{No-Intent} \emph{Health} personas and even higher prices for \emph{Intent} \emph{Health} personas. 
Similarly, bid ratios were significantly above the average across all personas for \emph{Health}, \emph{Kids} and \emph{Sports}. 
While \emph{No-Intent} \emph{Sports} personas attract significantly lower than average bids,  the  \emph{Intent} \emph{Sports} persona drew significantly higher bids than the average -- bidders are willing to pay higher prices for \emph{Intent} \emph{Sports} personas. 
Third, we note that the \emph{Intent} \emph{Science} persona attracts the lowest average bids and nearly the same as the \emph{No-Intent} \emph{Science} persona. 
Similarly, bid ratios were below the average across all categories for the \emph{Science}, \emph{Recreation} and \emph{Control} personas. 
Finally, we note significantly high variation in bid ratios across bidders for the \emph{Health} and \emph{Kids} personas, which also receive significantly higher bid ratios than other personas.  
Overall, variability in average bid ratios allows us to conclude that both \emph{user intent and persona impact bids placed by an advertiser.}

\begin{table}[!t]
\centering
  \begin{tabular}{lp{.35in}p{.35in}p{.35in}p{.35in}p{.35in}p{.35in}p{.25in}}
       & App. & Rub. & IX & OpX & Pub. & Avg. & Std.  \\	 
    \midrule
        Adult& 0.21\boldred{$^\downarrow$} & 0.43\boldred{$^\uparrow$} & 0.25 & 0.34 & 0.33 & 0.30 & 0.08 \\
        Arts& 0.34$^\uparrow$ & 0.45\boldred{$^\uparrow$} & 0.29\boldred{$^\downarrow$} & 0.37 & 0.36 & 0.36 & 0.05 \\
        Business& 0.28 & 0.45 & 0.28 & 0.30&0.51\boldred{$^\uparrow$} & 0.36 & 0.10 \\
        Computers& 0.20 & 0.75\boldred{$^\uparrow$} & 0.21 &  0.55 & 0.73\boldred{$^\uparrow$}$^\uparrow$ & 0.45\boldred{$^\uparrow$} & 0.25\boldred{$^\uparrow$} \\
        Games& 0.21 & 0.33\boldred{$^\uparrow$} & 0.21 & 0.34\boldred{$^\uparrow$} & 0.25$^\downarrow$ &  0.26\boldred{$^\downarrow$} & 0.06 \\
        Health& 0.21 & 1.16\boldred{$^\uparrow$}$^\uparrow$ & 0.28 &  0.94$^\uparrow$ & 0.54 & 0.59\boldred{$^\uparrow$} & 0.39\boldred{$^\uparrow$} \\
        Home& 0.20\boldred{$^\downarrow$} & 0.39\boldred{$^\uparrow$} & 0.24 & 0.28$^\downarrow$ & 0.28$^\downarrow$ & 0.27\boldred{$^\downarrow$} & 0.07 \\
        Kids& 0.20 & 0.41 & 0.18\boldred{$^\downarrow$}$^\downarrow$ & 0.47\boldred{$^\uparrow$} & 0.30 & 0.30 & 0.11 \\
        News& 0.19 & 0.61\boldred{$^\uparrow$} & 0.24 & 0.41 & 0.33 & 0.34 & 0.16 \\
        Recreation& 0.31$^\uparrow$ & 0.67\boldred{$^\uparrow$} & 0.23\boldred{$^\downarrow$} & 0.36 & 0.44 &  0.41 & 0.16 \\
        Reference& 0.18\boldred{$^\downarrow$} & 0.53 & 0.20 & 0.58\boldred{$^\uparrow$}$^\uparrow$ & 0.52 & 0.37 & 0.18 \\
        Regional& 0.23 & 0.43 & 0.33 & 0.73\boldred{$^\uparrow$}$^\uparrow$ & 0.35 & 0.39 & 0.17 \\
        Science& 0.30 & 0.70\boldred{$^\uparrow$} & 0.28\boldred{$^\downarrow$} & 0.44 & 0.58$^\uparrow$ & 0.46\boldred{$^\uparrow$} & 0.17 \\
        Shopping& 0.40\boldred{$^\downarrow$}$^\uparrow$ & 0.56 & 0.45$^\uparrow$ & 0.60\boldred{$^\uparrow$}$^\uparrow$ & 0.47 & 0.49\boldred{$^\uparrow$} & 0.07 \\
        Society& 0.22\boldred{$^\downarrow$} & 0.41 & 0.27 & 0.45\boldred{$^\uparrow$} & 0.37 & 0.32 & 0.09 \\
        Sports& 0.19 & 0.35\boldred{$^\uparrow$} & 0.13\boldred{$^\downarrow$}$^\downarrow$ & 0.23$^\downarrow$ & 0.30 & 0.23\boldred{$^\downarrow$} & 0.07 \\
        Control& 0.20\boldred{$^\downarrow$} & 0.26$^\downarrow$ & 0.28 & 0.44\boldred{$^\uparrow$} &0.37\boldred{$^\uparrow$} & 0.29 & 0.08 \\
        \midrule
        Avg.& 0.24$^\downarrow$ & 0.54$^\uparrow$ & 0.26 & 0.45 &  0.44 &  \boldred{0.36} & \textbf{0.39}  \\
        Std.& 0.06$^\downarrow$ & 0.21$^\uparrow$ & 0.07$^\downarrow$ & 0.17$^\uparrow$ & 0.14$^\uparrow$ & \boldred{0.08} & \textbf{0.13}  \\
  \end{tabular}
  \caption {Impact of user personas.
  HB median CPMs (USD) across our 16 personas and control persona
  for the top five bidders (AppNexus, Rubicon, IX, OpenX, and PubMatic) and associated weighted Avg. and Std. among categories and bidders. Bid prices exceeding $\pm\sigma$ among categories are denoted with \boldred{$^\uparrow$} or \boldred{$^\downarrow$}. Bid prices exceeding $\pm\sigma$ among bidders are denoted with $^\uparrow$ or $^\downarrow$. Avg. and Std. among \emph{persona} weighted averages are in \boldred{bold red}. Avg. and Std. among \emph{bidder} weighted averages are in \textbf{bold black}.}
  \label{tab:quantify:nointent}
  \postcaption
\end{table}
\begin{table}[!t]
\centering
  \begin{tabular}{lp{.35in}p{.35in}p{.35in}p{.35in}p{.35in}p{.35in}p{.25in}}
       & App. & Rub. & IX & OpX & Pub. & Avg. & Std. \\	 
    \midrule
    Adult& 0.97 & 0.97 & 2.10\boldred{$^\uparrow$} & 0.85 & 0.95 & 1.13 & 0.44 \\
    Arts& 1.04\boldred{$^\downarrow$} & 1.48\boldred{$^\uparrow$} & 1.45 & 0.97\boldred{$^\downarrow$} &  1.32 & 1.26 & 0.21 \\
    Business& 1.02 & 1.09 & 2.66\boldred{$^\uparrow$} & 1.01 & 0.84 & 1.32 & 0.66 \\
    Computers& 1.06 & 1.19 & 2.38\boldred{$^\uparrow$} & 1.18 & 0.71\boldred{$^\downarrow$}$^\downarrow$ & 1.30 & 0.55 \\
    Games& 1.20 & 1.83 & 1.81 & 1.06\boldred{$^\downarrow$} &  1.80$^\uparrow$ & 1.53 & 0.34 \\
    Health& 1.85$^\uparrow$ & 1.24 & 1.34 & 5.96\boldred{$^\uparrow$}$^\uparrow$ & 1.21 &  2.42\boldred{$^\uparrow$} & 1.92\boldred{$^\uparrow$} \\
    Home& 1.31 & 0.92\boldred{$^\downarrow$} & 1.50\boldred{$^\uparrow$} & 1.12 & 1.21 & 1.19 & 0.19 \\
    Kids& 1.31 & 1.51 & 6.00\boldred{$^\uparrow$}$^\uparrow$ & 0.76 &  1.49$^\uparrow$ & 2.34\boldred{$^\uparrow$} & 1.99\boldred{$^\uparrow$} \\
    News& 1.05 & 1.14 & 3.57\boldred{$^\uparrow$} & 1.05 &  0.95 & 1.62 & 1.06 \\
    Recreation& 1.76\boldred{$^\uparrow$}$^\uparrow$ & 1.09 & 1.04 & 1.08 &  0.86 & 1.15 & 0.29 \\
    Reference& 1.01 & 1.06 & 2.80\boldred{$^\uparrow$} & 0.70 & 0.60$^\downarrow$ &  1.26 & 0.82 \\
    Regional& 1.04 & 2.24\boldred{$^\uparrow$}$^\uparrow$ & 1.46 & 0.96 &  0.83 & 1.35 & 0.54 \\
    Science& 1.11 & 1.02 & 0.81\boldred{$^\downarrow$}$^\downarrow$ & 1.12\boldred{$^\uparrow$} & 0.92 & 0.99\boldred{$^\downarrow$} & 0.12\boldred{$^\downarrow$} \\
    Shopping& 1.18 & 1.42 & 1.55 & 1.52 &  1.00\boldred{$^\downarrow$} & 1.35 & 0.21 \\
    Society& 1.30 & 2.15$^\uparrow$ & 2.52\boldred{$^\uparrow$} & 0.76\boldred{$^\downarrow$} &   0.92 & 1.51 & 0.69 \\
    Sports& 1.13\boldred{$^\downarrow$} & 3.00$^\uparrow$ & 3.69\boldred{$^\uparrow$}$^\uparrow$ & 2.85$^\uparrow$ &  1.57$^\uparrow$ &  2.43\boldred{$^\uparrow$} & 0.94 \\
    Control& 0.87$^\downarrow$ & 1.32\boldred{$^\uparrow$} & 1.33\boldred{$^\uparrow$} & 0.60\boldred{$^\downarrow$} &  0.92 & 1.01\boldred{$^\downarrow$} & 0.28 \\
    \midrule
    Avg.& 1.19 & 1.45 & 2.33$^\uparrow$ & 1.07 & 1.40 & \boldred{1.48} & \textbf{1.49}  \\
    Std.& 0.25$^\downarrow$ & 0.54$^\downarrow$ & 1.35$^\uparrow$ & 0.31$^\downarrow$ & 1.26$^\uparrow$ & \boldred{0.45} & \textbf{0.74}  \\
  \end{tabular}
  \caption{Impact of showing intent. Cells indicate the ratio of median bid
  values for personas showing intent vs. personas showing no intent for the top five bidders (AppNexus, Rubicon, IX, OpenX, and PubMatic) and associated weighted Avg. and Std. Ratios exceeding $\pm\sigma$ among categories are denoted with \boldred{$^\uparrow$} or \boldred{$^\downarrow$}. Ratios exceeding $\pm\sigma$ among bidders are denoted with $^\uparrow$ or $^\downarrow$. Avg. and Std. among \emph{persona} weighted averages are in \boldred{bold red}. Avg. and Std. among \emph{bidder} weighted averages are in \textbf{bold black}.}
  \label{tab:quantify:intent}
  \postcaption
\end{table}

\presubsub
\subsubsection{How does bidding behavior vary across advertisers?}
\label{sub:quantify:advertisers}
\postsubsub
We now turn our focus to uncovering the differences in the behavior of different bidders. 
\Cref{tab:quantify:nointent} and \Cref{tab:quantify:intent} illustrate the impact of personas and intent on the bidding behaviors of the five most frequently observed bidders -- AppNexus, Rubicon, IX, OpenX, and PubMatic. 
We breakdown our analysis based on bidders responses to modified personas and intent. 

\paratitle{Bidder response to different personas
(\Cref{tab:quantify:nointent}).} 
First, Rubicon generally bids more per impression than any other advertiser (0.54 USD CPM 1.4x above the average bidder)  -- regardless of persona.
In fact, Rubicon bids the highest average values for 9 of the 16 personas. 
Second, the \emph{Health}, \emph{Shopping} and \emph{Computers} categories generally attract the most interest from all the bidders. 
We also find that some bidders show an aversion towards certain personas (\eg \emph{IX}  - \emph{Sports}, \emph{Kids}) bidding even less than they did for the control persona which had no history attached to it. 
Finally, we see that OpenX bids significantly more per impression than any other advertiser to place ads in front of our control persona (0.44 USD CPM). 
At a high-level, our results allow us to conclude that \emph{different bidders have preferences and aversions for different personas and only a few personas are universally preferred}.

\paratitle{Bidder response to demonstrated intent
(\Cref{tab:quantify:intent}).} First, we see that while all of our bidders generally had positive responses to intent.
The bid ratio for IX is significantly more than other bidders (1.6x more than the average). 
In fact, IX had the highest intent to no-intent ratio for 10 of our 16 personas. 
Conversely, PubMatic was found to be the least reactive to intent with their average bid value increasing only by 1.07x.
Second, some bidders had increases of nearly 6X in bid values when certain personas demonstrated intent. 
In particular, OpenX showed a 5.96x increase in their bid values when confronted with an intent \emph{Health} persona. 
%
%
Similarly, IX showed a 6.00x increase in their bid values when intent was demonstrated by the \emph{Kids} persona. 
%
%
Finally, looking at responses to our intent control persona, we see bid increases only for Rubicon and IX (1.32x and 1.33x increase) while the average showed only a 1.01x increase.
This shows that, in general, for many bidders, the knowledge of user personas dominates the decision to increase bid values and intent is only used to decide the magnitude of this increase. At a high-level, our results allows us to conclude that \emph{bidders rarely have similar responses to demonstrated user intent}.

\begin{table}[!t]
\centering
  \begin{tabular}{lp{.35in}p{.35in}p{.35in}p{.35in}p{.35in}p{.35in}p{.25in}}
       & App. & Rub. & IX & OpX & Pub. & Avg. & Std.  \\	 
    \midrule
        Adult& 2.28$^\uparrow$ & 0.86 & 2.87 & 0.72 &  5.94\boldred{$^\uparrow$} & 2.04 & 1.51 \\
        Arts& 1.19 & 0.86 & 1.08$^\downarrow$ & 0.94 & 10.8\boldred{$^\uparrow$}$^\uparrow$ & 1.77 & 2.61\boldred{$^\uparrow$} \\
        Business& 2.14\boldred{$^\uparrow$}$^\uparrow$ & 0.86\boldred{$^\downarrow$} & 1.76 & 1.84 &  - & 1.48 & 0.54\boldred{$^\downarrow$} \\
        Computers& 1.06\boldred{$^\downarrow$} & 2.81 & 2.72 & 0.34\boldred{$^\downarrow$} &  1.50$^\downarrow$ & 2.20 & 0.86 \\
        Games& 0.53\boldred{$^\downarrow$} & 1.92 & 1.26 & 4.48\boldred{$^\uparrow$} &  2.58\boldred{$^\uparrow$}$^\downarrow$ & 1.62 & 0.95 \\
        Health& 2.65$^\uparrow$ & 3.83$^\uparrow$ & 3.00$^\uparrow$ & 1.47\boldred{$^\downarrow$} &  9.76\boldred{$^\uparrow$}$^\uparrow$ & 3.99\boldred{$^\uparrow$} & 2.43\boldred{$^\uparrow$} \\
        Home& 0.62\boldred{$^\downarrow$} & 2.50 & 2.79 & - & - & 2.19 & 0.86 \\
        Kids& 2.84$^\uparrow$ & 0.86 & 0.30\boldred{$^\downarrow$}$^\downarrow$ &  2.02 & 5.94\boldred{$^\uparrow$} & 1.74 & 1.33 \\
        News& 0.50$^\downarrow$ & 2.92\boldred{$^\uparrow$}$^\uparrow$ & 1.17 &0.77 &  4.86\boldred{$^\uparrow$} &  1.64 & 1.18 \\
        Recreation& 0.45\boldred{$^\downarrow$}$^\downarrow$ & 0.86 & 2.71 & 4.02\boldred{$^\uparrow$} & - &  1.64 & 1.15 \\
        Reference& 1.09 & 0.86 & 1.39 &2.73\boldred{$^\uparrow$} &  3.56\boldred{$^\uparrow$} &  1.34 & 0.72 \\
        Regional& 0.91 & 1.56 & 0.81$^\downarrow$ & 7.73\boldred{$^\uparrow$}$^\uparrow$ &  - &  2.32 & 2.72\boldred{$^\uparrow$} \\
        Science& 2.08$^\uparrow$ & 3.56\boldred{$^\uparrow$}$^\uparrow$ & 3.09$^\uparrow$ & 1.60\boldred{$^\downarrow$} & - & 2.69\boldred{$^\uparrow$} & 0.86 \\
        Shopping& 0.45\boldred{$^\downarrow$}$^\downarrow$ & 1.58 & 4.87\boldred{$^\uparrow$}$^\uparrow$ & 0.88 &  5.94\boldred{$^\uparrow$} & 2.56 & 2.02 \\
        Society& 1.73 & 4.00\boldred{$^\uparrow$}$^\uparrow$ & 1.81 & 0.94\boldred{$^\downarrow$}$^\downarrow$ & 3.29\boldred{$^\uparrow$} & 2.25 & 0.96 \\
        Sports& 1.03 & 0.86 & 2.29\boldred{$^\uparrow$} & 0.28\boldred{$^\downarrow$} &  - &  1.12\boldred{$^\downarrow$} & 0.57\boldred{$^\downarrow$} \\
        Control& 0.62 & 0.86 & 2.75\boldred{$^\uparrow$} & 0.72 &  - &  1.38 & 0.94 \\

        \midrule
        Avg.& 1.27 & 1.78 & 2.01 & 2.25 & 6.22$^\uparrow$ & \boldred{2.00} & \textbf{2.71}  \\
        Std.& 0.76$^\downarrow$ & 1.08$^\downarrow$ & 0.91$^\downarrow$ & 2.23$^\uparrow$ &  3.06$^\uparrow$ & \boldred{0.66} & \textbf{1.61}  \\
  \end{tabular}
  \caption {Impact of user personas on winning bids.
  HB median CPMs (USD) across our 16 personas and control persona
  for the top five bidders (AppNexus, Rubicon, IX, OpenX, and PubMatic) and associated weighted Avg. and Std. among categories and bidders. Bid prices exceeding $\pm\sigma$ among categories are denoted with \boldred{$^\uparrow$} or \boldred{$^\downarrow$}. Bid prices exceeding $\pm\sigma$ among bidders are denoted with $^\uparrow$ or $^\downarrow$. Avg. and Std. among \emph{persona} weighted averages are in \boldred{bold red}. Avg. and Std. among \emph{bidder} weighted averages are in \textbf{bold black}.}
  \label{tab:quantify:winning_bids}
  \postcaption
\end{table}

\presubsub
\subsubsection{How much do advertisers pay to reach users?}
\label{sub:quantify:winning_bids}
\postsubsub
We now turn our attention to understanding \emph{the price advertisers actually pay to reach users}. %
To answer this question, we first examine the subset of winning bids (\ie highest bid value in the first-price HB auction) by the five most common bidders for \emph{No Intent} personas as shown in \Cref{tab:quantify:winning_bids}.
First, we note that on average bidders pay \$2.00 USD CPM across all personas in order to serve ads -- 5.5x the average of the corresponding bid price in \Cref{tab:quantify:winning_bids}. 
We see this trend across the board for different personas and bidders. 
We conclude that bidders have to pay substantially higher prices than their average bids to win the auctions.
Second, we note that the average winning bid in HB is 3.4x the average winning bid of RTB (Table XI - Only category column in \cite{OLEJNIK14NDSSRTB}).
There are two main explanations for this difference: (1) auction type (HB typically uses first-price auction and RTB typically uses second-price auction); and (2) bidding structure (HB uses a flattened model to issue bid requests and RTB uses a tiered/waterfall model where bid requests received at lower tiers are interpreted by bidders as bid ``left-overs'').
Third, we observe some similarities and differences in winning bid trends across personas for HB and RTB \cite{OLEJNIK14NDSSRTB}.
For the \emph{Health} persona, we observe above average winning bids in both RTB and HB. 
For other personas such as \emph{Games} and \emph{Sports}, we observe a shift from higher than average bids in RTB to lower than average bids in HB.  
This shift could be due to differences in bidder affinity caused by changing preferences among advertising partners \cite{URL_JOHNHANCOCKLIFEINSURANCE,URL_ACTIVISIONPREGNACYAPP} or time/location of measurements \cite{OLEJNIK14NDSSRTB,PAPADOPOULOS17IMCRTB}.

\presubsub
\subsubsection{How common is underbidding?} \label{sub:quantify:underbidding}
\postsubsub
During our bid collection process, we observed many bidders making \emph{zero} bids -- \ie a bid of \$0 USD CPM for the impression. 
There are several reasons for these bids. 
First, incorrect configurations of HB can lead to zero bids by advertisers. 
For example, price granularity is a setting made available to publishers which in essence can enforce a minimum bid value.
Any bid received below this value is rounded down to zero. 
Advertisers making bids without correctly accounting for this parameter will generate zero bids.
Second, and more interestingly, zero bidding is a form of underbidding -- \ie purposely making low bids with the motivation to gain access to user data (e.g., synced cookie \cite{Papadopoulos19cookiesynchronization, ESTRADA2019REGULATIONRTB}) associated with the impression rather than to win the auction. 
Although most exchanges which facilitate RTB auctions typically ban such behaviour and enforce mandatory minimum bidding participation, we find no such enforcement in HB auctions. 
Bids gathered from our experiments allow us to measure the frequency of zero bids, yet they do not let us convincingly distinguish whether they are due to misconfiguration or nefarious intent. 

\begin{table}[t]
\centering
  \begin{tabular}{p{.95in} >{\centering\arraybackslash}p{.70in}
    >{\centering\arraybackslash}p{.705in} >{\centering\arraybackslash}p{.70in}}
    \multirow{2}{*}{Bidder} & \multicolumn{3}{c}{Percentage of zero bids} \\
                            & No-intent personas  & Intent personas & Total\\
    \midrule
        PubMatic*&				68.75&			66.37&			67.70\\
        AppNexus&				0.32&			0.26&			0.29\\
        IX*&				        19.14&			6.19&			13.53\\
        Rubicon*&				3.54&			2.27&			2.94\\
        OpenX*&				    1.04&			0.23&			0.66\\
        Criteo*&				    5.83&			1.60&			3.75\\
        Aol&				    8.71&			9.17&			8.90\\
        Sovrn*&				    10.43&			3.87&			7.10\\
        Districtm&				0.00&			0.00&			0.00\\
        Conversant&				0.00&			0.00&			0.00\\
    \midrule
    Total (all bidders) &				        24.21&			19.44&			22.07\\
  \end{tabular}
  \caption{The percentage of zero bids made, for intent and no-intent
  personas, by our 10 most commonly observed bidders. Bidders are sorted by
  total number of bids placed. Bidders appended with'*' indicate that 
  zero bidding behavior for a bidder was significantly impacted by 
  \emph{Intent} personas.}
  \label{tab:quantify:zerobids}
  \postcaption
\end{table}

\Cref{tab:quantify:zerobids} shows the fraction of zero bids received for our intent and no-intent persona for each of the top 20 bidders observed. 
First, we note that zero bidding is a common occurrence and they make up over 22\% of all bids. 
It is noteworthy that for two of the bidders -- PubMatic and Innity -- most of their bids are zero bids and together account for a vast majority of all zero bids observed in our measurements. 
Frequent underbidding observed for these two bidders is indicative of the absence of minimum bidding performance requirements in HB through contract guarantees, which are often enforced in RTB auctions \cite{ESTRADA2019REGULATIONRTB}. 
Second, we assess whether a bidder's likelihood of placing zero bids is \textit{significantly} impacted by personas demonstrating intent to make a transaction.  
To this end, we apply the chi-square proportions test \cite{CAMPBELSTATSINMED2007} to compare percentages of zero bids placed by a bidder for \emph{No-Intent} and \emph{Intent} personas.
We see a statistically significant (4.77\%) decrease in the percentage of zero bids when a persona shows intent to make a transaction. 
This suggests that bidders are more motivated to make positive bids when the user communicates transaction intent. 
Finally, we can hypothesize a bidder's motivation to place zero bids by comparing percentage of zero bids between \emph{No-Intent} and \emph{Intent} personas.
We suspect that intentional underbidding will lead to a statistically significant difference in the prevalence of zero bids between \emph{Intent} and \emph{No-Intent} personas.
We observe a significant decrease in zero bids for PubMatic, IX, Rubicon, OpenX, Criteo, Sovrn, which leads us to suspect that zero bids are unlikely to be caused as a result of configuration errors.

\presec
\section{Inferring Tracker-Advertiser Relationships} \label{sec:inferring} 
\postsec
We showed that an advertiser's assessment of the value of a user (\ie bids) is highly dependent on the available information (\ie browsing history). 
To get relevant user information, advertisers (or DSPs bidding on behalf of advertisers) gather information from different trackers (or DMPs in general) through client-side or server-side  mechanisms. 
Inferring such data sharing relationships between different entities in the online advertising ecosystem is an important problem. 
It is challenging to infer such data sharing relationships, particularly at the server-side because they are not directly observable at the client-side.
Next, we present our approach to infer tracker-advertiser data sharing relationships at the client-side or the server-side.

\presub
\subsection{Proposed Approach}\label{sub:inferring:approach}
\postsub
Our approach, named \name, leverages the information provided by HB to infer data sharing relationships between trackers and advertisers. 
Our key insight is that an advertiser's bids for a persona will change when it has an information flow originating from \emph{some} tracker which has seen the persona before. 
This insight allows us to use bids, which are observable at client-side in HB, as a proxy for the existence of information flows (\ie data sharing relationship) between a tracker and a bidder. 
Thus, through careful manipulation of tracker exposure while constructing personas, we can analyze an advertiser's bids to make inferences about its data sharing relationships with the exposed trackers.

\begin{figure}[htbp] 
\precaption
\centering
\includegraphics[width=.85\columnwidth]{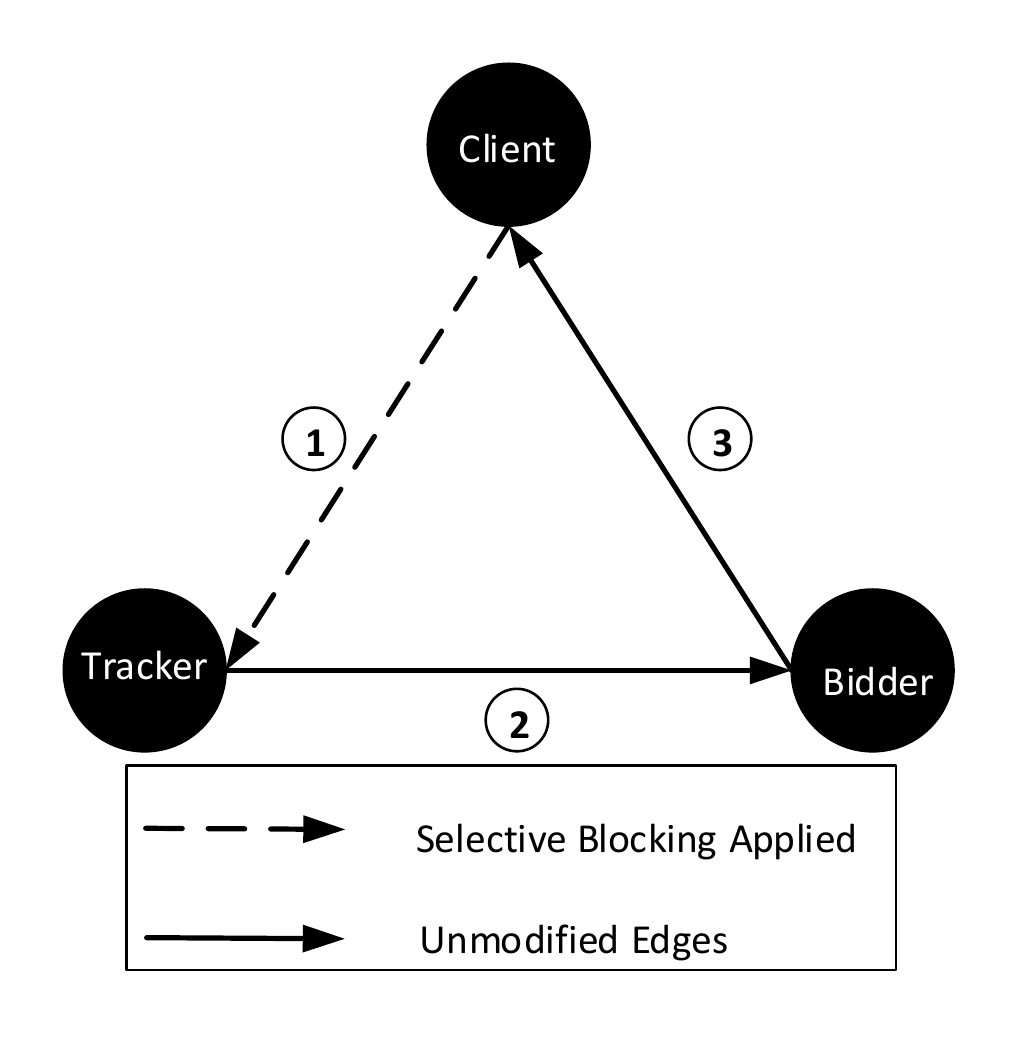}
\precaption
\caption{A simplified model of the information flows between trackers and bidders.}
\label{fig:TRIANGLE} 
\postcaption
\end{figure}

We illustrate \name with a simplified model showing information flows among key entities in the HB and the associated tracking ecosystem in \Cref{fig:TRIANGLE}. 
We start with edge \circled{1}, from the client to a tracker. 
This edge denotes data being gathered from clients by trackers partnering with publishers. 
Notice that these edges can be observed and manipulated at the client -- \ie trackers can be identified and blocked at the client and thus these edges can be deleted. 
Edge \circled{2} denotes the flow of data from trackers to advertisers. 
It is the presence of these flows that impacts the advertiser's bids for a user. 
Unfortunately, these edges, which are crucial for verifying regulatory compliance and building effective privacy-enhancing tools, are not observable or manipulated by the client. 
Finally, edge \circled{3} denotes the bid sent by an advertiser to a client in HB. 
Notice that these edges are observable by the client in HB. 
\emph{Our goal is to infer the existence of edge \circled{2} given the ability to observe and manipulate edges \circled{1} and \circled{3}}. 
We do this as follows:
\begin{itemize}
\item We gather a large number of bids from different advertisers (edge \circled{3}) while exposing client personas to a select set of trackers (by selectively deleting edge \circled{1}).
\item We then train a machine learning model to predict the bid values placed by each advertiser based on tracker presence/absence as features. 
We argue that a model that is able to accurately predict bid values also uncovers data sharing relationships between advertisers and trackers. 
An accurate model will evaluate tracker presence to gauge their impact on bid values.  
Put another way, if a machine learning model given edge \circled{1} as features is able to predict the values of edge \circled{3}, then it must have automatically inferred the presence or absence of edge \circled{2}.
\item Next, we analyze our interpretable machine learning models to identify the features (\ie edge \circled{1}) which had the most impact on our trained model's bid predictions. 
The information gain of these features establishes the likelihood of a relationship between the tracker and advertiser (\ie edge \circled{2}). 
Put another way, trackers that have a high impact on our model for a particular advertiser are more likely to have a data sharing relationship with the advertiser than those having no impact on our model. 
\emph{After all, under the assumption of one tracker per page, if deleting a tracker edge consistently has no impact on the bid values from the advertiser, it must be true that there is no relationship between the tracker and advertiser.}
As we discuss next, our method can generalize to multiple trackers.
\end{itemize}

\presub
\subsection{Measurement Method}\label{sub:inferring:methodology}
\postsub
\Cref{tab:inferring:questions} illustrates the contributions of this work towards inferring advertiser-tracker data sharing relationships.
To answer the questions listed in \Cref{tab:inferring:questions}, we conducted controlled measurements as follows. 
At a high-level, our method is explained by: (1) how we selectively expose trackers to information about our personas (how we manipulate edge \circled{1});  (2) how we measure the bids made by advertisers for each of our personas (how we observe edge \circled{3}); (3) how we predict bids; and (4) how we identify and validate influence of a tracker on an advertiser.
\begin{table}[!t]
  \centering
  \small
  \begin{tabular}{p{2.75in}c}
    Question & Results \\
    \midrule
    \textbf{Is tracker presence a predictor for advertiser bids?}
    & \Cref{subsubsection: 1 tracker predict bid price} \\
    \textbf{Which trackers influence the behaviors of which
    advertisers?}
    & \Cref{subsubsection: 2 tracker influence bidder behavior} \\
  \end{tabular}
  \caption{Questions answered by our study. Questions in \textbf{bold} have not
  been answered by previous work.}
  \label{tab:inferring:questions}
  \postcaption
\end{table}

\paratitle{Exposing user personas to trackers (manipulating edge \circled{1}).}
We constructed 10,000 user personas, each exposing selective characteristics to some subset of trackers, using the following approach. 

\begin{itemize}
\item \emph{Tracker exposure.} We used EasyList \cite{URL_EASYLIST} and EasyPrivacy \cite{URL_EASYPRIVACY} in combination with the outcome of the most recent Alexa top 1-million site crawl \cite{Englehardt16CCSOWPM} to obtain the top 20 most frequently observed tracking organizations and the tracking domains owned by them.\footnote{These 20 organizations are: Adobe,
Alibaba,
Alphabet,
AppNexus,
Automattic,
Baidu,
Comscore,
Criteo,
DoubleVerify,
ExoClick,
Facebook,
Integral Ad Science,
Microsoft,
Oracle,
PubMatic,
Quantcast,
Sovrn,
Twitter,
Verizon, and 
Yandex.}
We then randomly selected one organization and blocked all their trackers when building a user persona.
Specifically, trackers from the selected organization were blocked during the crawling of persona and intent sites described below. 
\item \emph{Persona selection.} We first randomly selected a persona which for the user persona to mimic. 
We used the same approach described in \Cref{sec:quantifying:methodology}, with the caveat that user persona was only constructed from a random subset of 1-10 of the Alexa top-50 sites within a persona category (rather than all of the top 50 sites). 
This reduction was necessary to scale of our experiments to build 10,000 user personas.
\item \emph{Intent selection.} Finally, we randomly assigned some of our
    personas to demonstrate intent to complete an online transaction. The
    method used was identical to the intent signaling mechanism described in
    \Cref{sec:quantifying:methodology}.
\end{itemize}
After building each user persona, we waited at least 90 minutes before moving into the bid collection phase.

\paratitle{Measuring advertiser bids (recording edge \circled{3}).} In order to measure the impact of selectively blocking edge \circled{1}, we needed to measure the values obtained through edge \circled{3}.
This was accomplished by visiting \emph{one} HB-enabled site using each trained persona and gathering the bids made by advertisers. 
We limited bid gathering to only one site since visiting multiple sites could result in tracker flows from the first site influencing the bids measured on subsequent sites.
Visiting a single site allows us to ensure that any tracker-advertiser information flow originates during the persona building phase and not during the bid collection.

\paratitle{Predicting bids.} Since our recorded bid values are continuous, we need a method to discretize them. 
Our bid values were discretized, in a similar manner as previous work seeking to predict encrypted bid values \cite{PAPADOPOULOS17IMCRTB}, by dividing bid values into classes. 
Specifically, we divided our dataset of bids values into three classes with the following bid ranges: 
$[-\infty, \mu - \sigma)$ \emph{low}, $[\mu - \sigma, \mu + \sigma]$ \emph{medium}, and $(\mu + \sigma, +\infty]$ \emph{high}, where $\mu$ is the mean bid value and $\sigma$ is the standard deviation of bid values. 
Next, we trained a separate Random Forest classifier for each advertiser with the goal of predicting bid classes given the presence of trackers as features. 
The Random Forest classifier was explicitly chosen due to its interpretable decision tree classification model. 
We applied 10-fold cross-validation to validate the accuracy of our constructed models. 
An accurate model for an advertiser demonstrates that tracker presence is a good predictor for bid class. 

\paratitle{Validating tracker influence on an advertiser.} 
We want to rank trackers based on their influence on advertiser generated bids. The decision trees produced by the Random Forest classifier rank features based on their importance. The most influential feature, with the highest information gain, is the root node of the tree. The subsequent nodes at lower levels have decreasing information gain on the partitioned data. Thus, given a reasonably accurate model of advertiser's bidding behavior, we analyzed decision trees to obtain a list of trackers ranked by their influence on each advertiser.  We then validated this list by comparing the observed relationships with the following sources of known tracker-advertiser relationships:
\begin{itemize}
    \item \emph{External databases.} We manually searched a variety of sources (\eg Crunchbase, public company websites, ad-tech blogs, \etc) to obtain publicly disclosed advertiser-tracker relationships.
    \item \emph{Client-side cookie syncing.} The entities in online advertising ecosystem use the \textit{cookie syncing} mechanism to share user identifiers at the client-side while circumventing the browser's same-origin policy. Using the heuristic presented in \cite{Papadopoulos19cookiesynchronization}, we detect client-side cookie syncing by looking for identifiers in the URL and the referrer field during our measurements. 
\end{itemize}

\presub
\subsection{Results} 

\subsubsection{Can tracker presence be used to predict bids?} 
\label{subsubsection: 1 tracker predict bid price}
\postsubsub
We now evaluate whether tracker presence can be used as predictors of advertiser bids. 
\Cref{tab:Results:Model_Performance} presents the classification performance of trained machine learning models for the top-5 bidders in our dataset. 
We note that trained machine learning models can predict bids by different bidders with reasonable accuracy.
Specifically, the accuracy ranges from 75\% for AppNexus to 83\% for IX.  
It is noteworthy that our trained machine learning models provide comparable accuracy to prior work on predicting encrypted bid values in RTB (82\%) \cite{PAPADOPOULOS17IMCRTB}. 
Thus, we conclude that our trained machine learning models can leverage \emph{tracker presence to accurately predict bids by different advertisers}.

\begin{table}[htbp]
\centering
  \begin{tabular}{p{.95in} >{\centering\arraybackslash}p{.70in}
    >{\centering\arraybackslash}p{.705in} >{\centering\arraybackslash}p{.70in}}
    Bidders                 & Accuracy   \\
    \midrule
    AppNexus                & 75\%    \\
    IX	                    & 83\%   \\
    Openx	                & 81\%    \\
    Rubicon	                & 82\%   \\
    PubMatic                & 78\% \\
    Avg. 	                & 80\% \\
  \end{tabular}
  \caption{Bid prediction accuracy of machine learning models for top-5 bidders in our dataset.}
  \label{tab:Results:Model_Performance}
  \postcaption
    \vspace{.05in}
\end{table}

\presubsub
\subsubsection{Tracker Influence On Bidder Behavior}
\label{subsubsection: 2 tracker influence bidder behavior}
\postsubsub

To understand a tracker's influence on an advertiser's bidding behavior, we use the decision tree model generated by our machine learning classifier. 
As discussed earlier, trackers at the top of the decision tree are more influential than those at the bottom. 
\Cref{tab:validating_tracker_Advertiser} lists the top-3 trackers for each of the top-5 bidders in our dataset. 
We note that different trackers are the most influential across different bidders. 
For example, our model ranks DoubleVerify as the most influential tracker for AppNexus while Alphabet as the most influential tracker for PubMatic. 
We observe that 11 of 15 advertiser-tracker relationships inferred by our model are validated by external databases (10 of 15) or client-side cookie syncing (4 of 15).
3 of these advertiser-tracker relationships are validated by both external databases and client-side cookie syncing.
We note 11 potential server-side advertiser-tracker relationships that are not validated using client-side cookie syncing.
Of these 11, we are able to validate 7 such server-side relationships using external databases. 
The remaining 4 may be attributed to previously unknown server-side data sharing relationships, imperfect heuristics to detect cookie syncing, or erroneous inferences by \name.

\presubsub
\subsubsection{Implications}
\label{subsubsection: 3 Implications}
\postsubsub
It is noteworthy that \name is able to uncover several server-side advertiser-tracker relationships that are not observable at the client-side.
Our findings seem to indicate that online advertising and tracking ecosystems may be shifting from the client-side to the server-side. 
We argue that there are several motivations for such a shift from client-side to server-side. 
First, and perhaps most importantly, advertisers and trackers are shifting to server-side data sharing to circumvent client-side blocking tools. 
Specifically, a significant fraction of users have installed browser extensions (\eg uBlock Origin \cite{URL_UBLOCKORIGIN}, Adblock Plus \cite{URL_ABP}, Ghostery \cite{URL_GHOSTERY}, Privacy Badger \cite{URL_PRIVACY_BADGER}, \etc) to block ads and trackers at the client-side. 
Moreover, mainstream browsers such as Safari and Firefox have enabled anti-tracking protections by default \cite{URL_ITP_20,URL_FF_ETP}. 
We believe that advertisers and trackers are likely shifting to server-side data sharing to circumvent client-side blocking mechanisms. 
Second, advertisers and trackers also prefer server-side implementations due to performance reasons. 
Specifically, client-side implementation of resource-heavy advertising and tracking logic significantly degrades page load performance \cite{URL_IPHONE_TRACKING_APPS,URL_S2SAPPNEXUS}.
Moreover, client-side implementations are also susceptible to slow response times resulting in auction timeouts (bids arriving after an auction timeout occurs are ignored) \cite{URL_S2SAPPNEXUS}.
To conclude, our findings highlight the shift from client-side to server-side data sharing in the online advertising ecosystem. 
As server-side data sharing---which can be inferred by \name---becomes more prevalent, it is unclear whether the current generation of client-side blocking tools would continue to remain effective.

\begin{table}[!t]
    \vspace{.05in}
  \centering
  \small
    \begin{tabular}{c c c c}    
     & Tracker 1 & Tracker 2 & Tracker 3 \\
    \midrule
    AppNexus & DoubleVerify \textsuperscript{\cite{URL_VALIDATION_APPNEXUS_DOUBLEVERIFY}} & 
    Automattic \textsuperscript{\cite{URL_VALIDATION_AUTOMATIC_APPNEXUS}} &
    Comscore \textsuperscript{{CS,{\cite{URL_VALIDATION_COMSCORE_APPNEXUS}}}} \\
    IX & Sovrn \textsuperscript{\cite{URL_VALIDATION_APPNEXUS_DOUBLEVERIFY}} & 
   PubMatic \textsuperscript{\cite{URL_IX_SOVRN_PUBMATIC}} &
    DoubleVerify \textsuperscript{\cite{URL_IX_SOVRN_PUBMATIC}} \\
    OpenX & Microsoft \textsuperscript{\cite{URL_VALIDATION_OPENX_MICROSOFT}} &
    AppNexus \textsuperscript{{CS,\cite{URL_VALIDATION_OPENX_APPNEXUS}}} &
    Criteo \textsuperscript{{\cite{URL_VALIDATION_OPENX_CRITEO}}} \\
    Rubicon & Verizon  \textsuperscript{CS,{\cite{URL_VALIDATION_RUBICON_VERIZON_2}}} & DoubleVerify &  Facebook \\
    PubMatic & Alphabet \textsuperscript{CS} &    Twitter & Microsoft  \\
  \end{tabular}
  \caption{Tracker influence is ranked in the descending order of information gain for each of the top-5 bidders in our data set. The bidder-tracker relationships that we are able to validate using manual search are marked with a citation. Cookie syncing detected using client-side analysis are marked with $^{CS}$.} 
  \label{tab:validating_tracker_Advertiser}
  \postcaption
  \vspace{-.05in}
\end{table}
\presub
\subsection{Limitations}
\postsub

\paratitle{Completeness issues.} 
Our study makes two simplifying assumptions that may impact the completeness of our results. 
    First, we restrict our inferences to only include bidder relationships with the top-20 tracking organizations. As a result, we are unable to draw inferences about bidder relationships with smaller tracking services (rank $>$ 20). We argue that, given the extreme skew in tracker coverage across the web \cite{Englehardt16CCSOWPM}, our approach would capture the overwhelming majority of data sharing occurring in the advertising and tracking ecosystem. 
    Second, the data sharing relationships in the online advertising ecosystem may be indirect. More specifically, trackers may share data with many non-bidder entities (e.g., SSP, AdX) and bidders may gather data from different data sources (DMP). Our approach is unable to determine whether a tracker-bidder relationship is direct or indirect (\ie involves other intermediaries). However, as long as the presence of a tracker impacts the bids, our approach is able to infer that there is a direct or indirect tracker-bidder relationship. 
    
\paratitle{Correctness issues.} Our study also makes several simplifying assumptions that may impact the correctness of our results.
  First, the accuracy of our machine learning approach is not perfect. 
    It is possible that some of the tracker-bidder relationship inferences based on our trained machine learning models are incorrect. 
    To overcome this limitation, we take a conservative approach by limiting ourselves to top-3 trackers identified by our machine learning models.
    As part of our future work, we plan to investigate automated methods to determine the optimal cutoff point given a certain error tolerance. 
  Second, our approach may fail in the presence of tracker-tracker data sharing relationships.
    Consider an example where the following data sharing relationships are observed: ($T_1$,
    $T_2$), ($T_2$, $A$) where $T_1$ and $T_2$ are trackers and $A$ is an
    advertiser. 
    Our technique might conclude that there is no relationship
    between $T_2$ and $A$ as a consequence of not observing a change in bidding
    behavior from $A$ when blocking $T_2$. 
    However, this conclusion might be
    incorrect if the reason for no change in $A$'s behavior is the flow of
    information from $T_1$ to $A$ via $T_2$. 
    We mitigate this problem in our work by analyzing trackers at the organizational level. 
    In other words, we assume that all domains within an organization (\eg doubleclick.net and google-analytics.com belong to Alphabet) \emph{will}
    share data with each other.

\presec 
\section{Related Work} 
\label{sec: related work} 
\postsec
Prior work related to our research can be categorized into two types: (1) user value quantification and (2) characterization of entities and their relationships in the online advertising and tracking ecosystem.
\presub 
\subsection{Quantifying the Value of a User} 
\postsub
As more advertisers rely on online advertising \cite{URL_CNBCGLOBALADSPEND} and more Internet users have the expectation of free services \cite{Dou-jar2005, Lin-eComm2013, Wang-eComm2005}, online behavioral advertising, facilitated by tracker gathered user data, has become the dominant monetization model on the web.
Much of prior work has sought to uncover the value of different types of users (and their data) to different entities in the online advertising and tracking ecosystem.
Along these lines, there has been a great deal of interest in understanding how much value users place in the data that they trade for free access to online services.
These studies have generally borrowed techniques from psychology and economics to design experiments to implicitly uncover the value that users place on their data.
Findings have shown that context dictates privacy valuations of data \cite{Acquisti-jls2013,Grossklags-weis2007,Huberman-sp2005}, trustworthiness and intention of the buyer plays a role in privacy valuations \cite{Cvrcek-wpes2006, Danezis-weis2006}, and there is a mismatch in the actual and perceived value of user data \cite{Carrascal2013WWWBIGMAC, GONZALEZACMCHI17CHIFDTV, PAPADOPOULOS18WWWCOSTAD}.
From another perspective, there have been efforts \cite{OLEJNIK14NDSSRTB, PAPADOPOULOS17IMCRTB, PAPADOPOULOS18WWWCOSTAD} to quantify how much user data is worth to advertisers.
Such efforts are generally more challenging due to the opacity of the advertising ecosystem -- \ie it is difficult to uncover exactly how much advertisers are paying (bidding) to place ads in front of specific users.
These works leveraged the visibility afforded to the user's browser in the RTB auction to uncover the winning bids.
More specifically, these works leveraged the fact that the winning bid notification in an RTB auction (including information about the winner and the winning bid value) is relayed to the browser in step \circled{9} of the RTB workflow shown in Figure \ref{fig:RTBHB}.

In a seminal work, Olejnik \etal \cite{OLEJNIK14NDSSRTB} analyzed RTB winning bid notifications to understand variation across different user personas based on their location, time, and browsing history.
The authors reported that advertisers pay as little as \$0.0005 per site visit.
Further, they showed that the prices that advertisers pay vary based on browsing histories reflecting different generic interests (\eg, games, news, shopping) and specific intents (\eg, hotel booking, jewelry, electronics).
Our work differs and builds upon this work in the following ways.
\begin{itemize}
  \item First, we are able to shed light on the bidding behavior of different
    advertisers as we are able to capture bids from each advertiser, not just the
    winning bid.
  \item Second, they encountered and ignored encrypted bids, which were
    (incorrectly \cite{PAPADOPOULOS17IMCRTB}) assumed to be comparable to plain
    text bids.
In contrast, we do not encounter encrypted bids in our HB  measurements.
  \item Third, because we can observe bids from different advertisers in
    HB, we are able to show that advertisers bid differently
    (by as much as 5.5x for \textit{Health} category in \Cref{tab:quantify:intent}) for the same user personas.
  \item Further, through controlled experiments in the second phase of our
    study, we are able to show that bid variations across different advertisers
    are, in part, due to differences in advertiser-tracker relationships.
\end{itemize}

In a follow-up work, Papadopoulos \etal \cite{PAPADOPOULOS17IMCRTB} addressed the limitation placed by encrypted bid values (encountered by \cite{OLEJNIK14NDSSRTB}) by developing a machine learning approach to infer values of encrypted winning bids with 82\% accuracy.
The authors showed that encrypted bids are are $1.7$X higher than bids sent in the clear.
We build on this work by seeking to understand tracker-advertiser relationships using a similar machine learning approach for modeling bid values.
However, unlike their work, we are not interested in predicting encrypted bid values because we do not encounter encrypted bids in HB.
Instead, we leverage a machine learning model that can accurately predict an advertiser's bids based on information about presence/absence of trackers to infer tracker-advertiser relationships.
\presub
\subsection{Characterization of Advertising and Tracking Entities and Relationships} 
\postsub
There have been many studies which have sought to measure the prevalence of different entities in the advertising and tracking ecosystem.
These include large-scale and longitudinal crawls measuring the prevalence of trackers and different tracking techniques on the web \cite{Englehardt16CCSOWPM,Merzdovnik17BlockMeIfYouCanESP,MAYER2012IEEESPFOURTHPARTY,lerner16internetjones,Papadopoulos19cookiesynchronization}, mobile \cite{Razaghpanah-ndss2018, Rodriguez-imc2012, Le-c2bid2015, Reyes-conpro2017, Backes-ccs2016, Chen-hotmobile2014}, and across multiple platforms \cite{Zimmeck-security2017, Brookman-pets2017}.

Notably, Englehardt and Narayanan \cite{Englehardt16CCSOWPM} studied the prevalence of different stateful and stateless techniques on the Alexa top 1-million websites.
They reported that a few third-parties including Google, Facebook, Twitter,Amazon, and AdNexus cover at least 10\% of the top 1M websites.
They also showed that client-side cookie syncing is prevalent among third-parties: 45 of the top 50 third-parties sync cookies with at least one other party and the most popular third-party (\texttt{doubleclick.net}, an Alphabet-owned AdX) syncs cookies with 118 different third-parties.
This highlighted that trackers,  even when owned by different organizations, often exchanged data to improve  participation in online advertising.
In addition to information flows among trackers that are observable at the client-side (\ie cookie syncing), researchers have also investigated methods to detect server-to-server (S2S) information flows among trackers.
This is much more challenging since they are not observable at the client-side (browser).

To address this challenge, Bashir \etal conducted controlled experiments and inferred a small subset of S2S information flows by investigating the process of ad {retargeting} \cite{Bashir16USENIXTRACING}.
Since retargeting necessitates data exchange between two AdXes, the authors conducted controlled experiments to trigger and detect ad retargeting and infer S2S information flows.
The underlying intuition was that if a retargeted ad was served by an entity that did not observe the original visit which triggered the retargeted ad, then it must have got information about this visit through an S2S information flow.
We leverage the same underlying intuition as in Bashir \etal~\cite{Bashir18PETSDIFUSSION, Bashir16USENIXTRACING} to infer tracker-advertiser relationships.
However, instead of relying on retargeting as the ``signal'' for data exchange, we operationalize this intuition using a machine learning model that is trained to predict an advertiser's bid values using presence/absence information of popular trackers.

\presec
\section{Concluding Remarks} 
\label{sec:concluding_discussion}
\postsec
In this paper, we leveraged header bidding (HB) to gain insights into the bidding behavior of advertisers and presented a machine learning approach to infer data sharing relationships between advertisers and trackers. 
Our work advances the field along two main avenues. 
First, we are able to provide more nuanced insights into the bidding behavior of online advertisers.
While prior research \cite{OLEJNIK14NDSSRTB,PAPADOPOULOS17IMCRTB} was limited to analyzing only the winning bids in RTB, we are able to observe all bids made by different advertisers in HB. 
Second, we are able to infer data sharing relationships between advertisers and trackers irrespective of whether it is happening at the client-side or the server-side. 
While prior work could only detect client-side data sharing \cite{Papadopoulos19cookiesynchronization} or infer server-side data sharing relationships when retargeting occurs \cite{Bashir16USENIXTRACING}, our approach is able to infer client-side and server-side data sharing for any advertiser placing bids without relying on specific triggers such as retargeting.

Our work can help existing privacy-enhancing tools in presenting empirically derived inferences about (1) how the data is shared between entities in the online advertising and tracking ecosystems; and (2) what is the perceived value of users. 
Along the first direction, privacy enhancing tools such as Mozilla Lightbeam \cite{URL_LIGHTBEAM}, uBlock Origin \cite{URL_UBLOCKORIGIN}, and Ghostery \cite{URL_GHOSTERY} provide users transparency and control over online tracking. 
Our work can be used to address known limitations \cite{Bashir18PETSDIFUSSION} of these tools by identifying server-side data sharing practices of online trackers.
Along the second direction, our HB measurements can be used to improve existing user valuation tools such as \emph{RTBAnalyzer} \cite{OLEJNIK14NDSSRTB} and \emph{YourADValue} \cite{PAPADOPOULOS17IMCRTB} by capturing a more complete picture of \textit{all} advertisers' bidding behaviors. 
These measurements can further inform micropayment-based alternate web monetization models \cite{LESK2018IEEESP} (\eg Flattr \cite{URL_FLATTR}, Contributor \cite{URL_GOOGLECONTRIBUTOR}, BAT \cite{URL_BAT}) by suggesting how much users should pay a publisher in exchange for blocking ads.


\presec
\section*{Acknowledgement}
\postsec
This research is supported in part by a Google Faculty Research Award, a grant from the Data Transparency Lab, and the National Science Foundation under grant numbers 1715152 and 1815131.

\bibliographystyle{abbrv}
\bibliography{hb}
\end{document}